\documentclass[1p]{elsarticle}
\usepackage{amssymb}

\def\ii{\'{\i }}
\usepackage{amssymb}
\begin{document}
\begin{frontmatter}
\title{Elliptic flow: pseudorapidity and number of participants
dependence}
\author{ I. Bautista$^{1}$, J. Dias de Deus$^{2}$, C. Pajares $^{1}$}
\address{ IGFAE and Departamento de F\'\i sica de Part\'\i culas,
Univ. of Santiago de Compostela, 15706, Santiago de Compostela,
Spain, CENTRA, $^{2}$Departamento de F\ii sica, IST, Av. Rovisco
Pais, 1049-001 Lisboa, Portugal}
\begin{abstract}
We discuss the elliptic flow dependence on pseudorapidity and
number of participating nucleons in the framework of string
percolation, and argue that the geometry of the initial overlap
region of interaction, projected in the impact parameter plane,
determines the experimentally measured azimuthal asymmetries. We
found good agreement with data.
\end{abstract}
\end{frontmatter}
The discovery of the large elliptic flow $v_{2}$ was one of the
most important achievements at RHIC experiments [1-9]. A non
vanishing anisotropic flow exist only if the particles measured in
the final state depend not only on the local physical conditions
realized at the production but as well on the global event
geometry. In a relativistic local theory, this non local
information can only emerge as a collective effect, requiring
interactions between the relevant degrees of freedom, localized at
different points of the collision region. In this sense,
anisotropic flow is particularly unambiguous and convincing
manifestation of collective dynamics in heavy ion collisions [10].

The elliptic flow $v_{2}$ can be qualitatively explained as
follows. In a collision at high energy the spectators are fastly
moving opening the way, leaving behind at mid- rapidity an almond
shaped azimuthally asymmetric region of QCD matter. This spatial
asymmetry implies unequal pressure gradients in the transverse
plane, with a larger density gradient in the reaction plane ( in-
plane). As a consequence of subsequent multiple interactions
between degrees of freedom this spatial asymmetry leads to an
anisotropy in momentum plane. The final particle transverse
momentum is more likely to be in- plane than in the out- plane,
with $v_{2}>0$ as predicted [11].

The basic idea of our model [12] is that the angular azimuthal
anisotropy associated to the geometry of the first stages in the
collision - the projected almond- influences in a determinant way
the presence or not of the flow. In other words if the projected
overlap region was a circle we would have $v_{2}\equiv0$. As in
the almond case the small axis is in the reaction plane,
corresponding to higher matter density, then $v_{2}>0$.

Our model was introduced in [12] and a discussion of applications
and conjectures were presented dependence of $v_{2}$ on the
produced hadron, validity of quark counting rules, applications to
nuclear reduction factors, etc. Here we just want to call
attention to $v_{2}$ as a function of pseudorapidity $\eta$ and
the number of participants $N_{A}$ for nucleus A, after the
integration over $p_{T}^{2}$. The triangle shape shown by the data
on the dependence of $v_{2}$ on $\eta$ it is not easy reproduced
by models as it has been recently emphasized [13]. We show that
string percolation model is able to do it.

The string percolation model [14] develops around the concept of
transverse density $\eta^{t}$,
\begin{equation}
\eta^{t}=(\frac{r_{0}}{R})^{2}\bar{N}^{s}
\end{equation},
where $\bar{N}^{s}$ is the number of longitudinal strings formed
in the collision, $r_{0}$ is the radius of the single string and
$R$ the effective radius of the interaction overlap region $S$ in
the impact parameter b,
\begin{equation}
S=\pi R^{2}
\end{equation}
with
\begin{equation}
S=2R_{A}^{2}[cos^{-1}(\beta)-\beta\sqrt{1-\beta^{2}}]
\end{equation},
$R_{A}$ being the nuclear radius and
\begin{equation}
\beta=\frac{b}{2R_{A}}
\end{equation}

Two relations, one for the particle density $dn/d\eta$ and the other for
the average transverse momentum squared define the essential features of
the model [13,14]
\begin{equation}
dn/d\eta=F(\eta^{t})\bar{N}^{s}\mu_{1}
\end{equation}
and
\begin{equation}
<p_{T}^{2}>=<p_{T}^{2}>_{1}/F(\eta^{t})
\end{equation}
where $\mu_{1}$ and $<p_{T}^{2}>_{1}$ are single string parameters
and $F(\eta^{t})$ is the colour reduction factor [15]:
\begin{equation}
F(\eta^{t})=\sqrt{\frac{1-e^{-\eta^{t}}}{\eta^{t}}}.
\end{equation}
We introduce now two reasonable approximations: that $N^{s}$ is
proportional to the number of binary collisions and that $R$ is
proportional to the proton radius,
\begin{equation}
\bar{N}^{s}\sim\bar{N}^{s}_{p}N_{A}^{4/3}
\end{equation} and
\begin{equation}
R= R_{p}N_{A}^{1/3},
\end{equation}
where $N_{p}^{s}$ and $R_{p}$ are proton parameters and $N_{A}$ is
the number of participants from nucleus A.

From (1), (8) and (9) we obtain
\begin{equation}
\eta^{t}_{N_{A}}=\eta^{t}_{p}N_{A}^{2/3}.
\end{equation}

By using (9) and (1) on (5) one obtains
\begin{equation}
\frac{1}{N_{A}^{2/3}}\frac{dn}{d\eta}=F(\eta^{t})\eta^{t}(\frac{R_{p}}{r_{0}})^{2}\mu_{1}
\end{equation},
and we observe that the right hand side of (11) and (6) are deeply
related.

This kind of results appears in the Color Glass Condensate (CGC)
[16] and in string percolation [14,17]. Note that (11) can be
written in the form
\begin{equation}
\sqrt{(1- \exp{-\eta^{t}})\eta^{t}}=\frac{\pi
r_{0}^{2}}{\mu_{1}}[\frac{1}{S}\frac{dn}{d\eta}]
\end{equation}

This relation, as we shall see,
is essential to understand the (pseudo)rapidity and number of
participants per nucleus dependence of $v_{2}$: $v_{2}(\eta,
N_{A})$. Note that in (12) small $\eta^{t}$ corresponds to large
$\eta$ and large $\eta^{t}$ to small $\eta$.

Regarding $p_{T}^{2}$ distributions, we started with Schwinger
gaussian formula, including fusion and percolation (via
$F(\eta^{t})$) and clustering fluctuations (via the parameter
$k(\eta^{t})$) to obtain[18]:
\begin{equation}
\frac{d^{2}n}{dp_{T}^{2}d\eta}=
\frac{dn}{d\eta}\frac{k-1}{k}\frac{F(\eta^{t})}{<p_{T}^{2}>_{1}}\frac{1}{(1+\frac{F(\eta^{t})p_{T}^{2}}{k<p_{T}^{2}>_{1}})^{k}}.
\end{equation}
Most of the RHIC data are well described by formula (13)
[12,18,19].

In order to discuss directional production along the azimuthal angle
$\varphi$, we shall introduce a convenient variable
\begin{equation}
X=F(\eta^{t})p_{T}^{2},
\end{equation}
and $X_{\varphi}$
\begin{equation}
X_{\varphi}=F(\eta_{\varphi}^{t})p_{T}^{2},
\end{equation}
with
\begin{equation}
\eta_{\varphi}^{t}=\eta^{t}(\frac{R}{R_{\varphi}})^{2}
\end{equation}
such that we can simplify notation
\begin{equation}
\frac{dn}{dp_{T}^{2}dy}\rightarrow f(X)
\end{equation}
\begin{equation}
\frac{dn}{dp_{T}^{2}dyd \varphi}\rightarrow f(X_{\varphi})
\end{equation}
Expanding now $X_{\varphi}$ or $R_{\varphi}$ around $X$ or $R^{2}$
we write
\begin{equation}
f(X_{\varphi})\simeq \frac{2}{\pi} f(X)[\frac{1+ \partial ln
f(X)}{\partial R}(R^{2}_{\varphi}-R^{2})]
\end{equation}
Note that (12) satisfies the normalization condition
\begin{equation}
\int_{0}^{\pi /2} f(X_{\varphi})d \varphi =f(X)
\end{equation}
because $R^{2}=<R_{\varphi}^{2}>[12]$. Finally we obtain for
$v_{2}$, a function of several variables including $p_{T}^{2}$,
$\eta$ and $N_{A}$,
\begin{equation}
v_{2}=\frac{2}{\pi}\int_{0}^{\pi/2} d \varphi cos(2\varphi)
(\frac{R_{\varphi}}{R}^{2})\frac{1}{2}\frac{e^{-\eta^{t}}-F(\eta^{t})^{2}}{F(\eta^{t})^{2}}\frac{
F(\eta^{t})p_{T}^{2}/<p_{T}^{2}>_{1}}{1+F(\eta^{t})^{2}p_{T}^{2}/k<p_{T}^{2}>_{1}}
\end{equation}
which we shall write as the product of tree factors,
\begin{equation}
v_{2}=[\varphi][\eta^{t}][F(\eta^{t})p_{T}^{2}],
\end{equation}
\begin{equation}
[\varphi]=\frac{2}{\pi}\int_{0}^{\pi/2}d\varphi
cos(2\varphi)(\frac{R_{\varphi}}{R}^{2}),
\end{equation}
or, having present that
\begin{equation}
\frac{R_{\varphi}}{R_{A}}=\frac{\sin{\varphi-\alpha}}{\sin{\varphi}}
\end{equation}
and
\begin{equation}
\alpha=sin^{-1}(\beta sin \varphi),
\end{equation}
\begin{equation}
[\varphi]=\frac{2}{\pi}\int_{0}^{\pi/2} d\varphi
\frac{cos(2\varphi)sin^{2}(\varphi-\alpha)}{sin^{2}\varphi}(\frac{R_{A}}{R})^{2},
\end{equation}
\begin{equation}
[\eta^{t}]=\frac{1}{2}\frac{e^{\eta^{t}}-F(\eta^{t})^{2}}{F(\eta^{t})^{2}}
\end{equation}
and
\begin{equation}
[F(\eta^{t})p_{T}^{2}]=\frac{F(\eta^{t})p_{T}^{2}/<p_{T}^{2}>_{1}}{1+\frac{F(\eta^{t})p_{T}^{2}}{k
<p_{T}^{2}>_{1}}}.
\end{equation}
Let us next look at the factor $[\varphi]$, (23) in (21) and
consider, for fixed $\eta$ and $\sqrt{s}$, two limits:

 $i)$ $b\rightarrow 0$ or $\beta\rightarrow 0$ or  $N_{A} \rightarrow A$
which implies, (25), $\alpha \rightarrow 0$. We then have
$[\varphi]\rightarrow 0$, $[\eta^{t}]$, (27),$\rightarrow $
constant, and $[F(\eta^{t})p_{T}^{2}]\rightarrow 0 $, or:
\begin{equation}
N_{A}\rightarrow A \mbox{ ,  } v_{2}(p_{T}^{2})\rightarrow 0
\end{equation}.

$ii)$ $b\rightarrow 2R_{A}$ or $\beta \rightarrow 1$ or
$N_{A}\rightarrow 0$ which implies, (18),
 $\alpha \rightarrow \varphi$. We then have $[\varphi]\rightarrow 0$, $[\eta^{t}]\rightarrow$ constant,
  $[F(\eta^{t})p_{T}^{2}]\rightarrow $ some finite function of $p_{T}^{2}$, or:
\begin{equation}
N_{A}\rightarrow 0 \mbox{ ,  } v_{2}(p_{T}^{2})\rightarrow 0
\end{equation}
If we look now to the $p_{T}^{2}$ dependence of $v_{2}$ in
$[F(\eta^{t})p_{T}^{2}]$, (28) we see that
 $v_{2}\rightarrow 0 $ as $p_{T}^{2}\rightarrow 0$
  and $v_{2}\rightarrow k \sim$ constant, as $p_{T}^{2}\rightarrow\infty$. This is observed in data [20].

We perform next the integration in $p_{T}^{2}$, weighted by
$\frac{dn}{dp_{T}^{2}d\eta}/\frac{dn}{d\eta}$, to obtain:
\begin{equation}
v_{2}=\frac{2}{\pi}\int_{0}^{\pi /2}d \varphi
cos(2\varphi)(\frac{R_{\varphi}}{R})^{2}\frac{e^{-\eta^{t}}-F(\eta^{t})^{2}}{2F(\eta^{t})^{3}}\frac{R}{R-1}=[\varphi][\eta^{t}]'
\end{equation},
$[\eta^{t}]'$ being different from $[\eta^{t}]$ and $\eta^{t}$
being related to $\eta$ by relation (12). Applying now to
$v_{2}(\eta)$ the arguments used for $v_{2}(p_{T}^{2})$ we have:
\begin{equation}
i) N_{A}\rightarrow A \mbox{,  } v_{2}(\eta)\rightarrow 0,
\end{equation}
with $[\eta^{t}]'$ being some negative number depending on $\eta$;
\begin{equation}
ii) N_{A}\rightarrow 0 \mbox{,  }v_{2}(\eta)\rightarrow 0,
\end{equation}
with $[\eta^{t}]'\rightarrow 0$.

In conclusion, both
$v_{2}(p_{T}^{2})$ and $v_{2}(\eta)$ go to zero as
$N_{A}\rightarrow A$ and $N_{A}\rightarrow 0$. In the case of
$v_{2}(p_{T}^{2})$ with $p_{T}^{2}\equiv0$, $v_{2}(p_{T}^{2})$ is
identically zero.

As $[\eta^{t}]$ is, in modulus, a growing function of $\eta^{t}$,
it is clear that $v_{2}(\eta)$, at fixed $\eta$, is a growing
function of energy and of $N_{A}$ see [20].

Regarding $v_{2}$ normalized by the eccentricity $\epsilon$,
\begin{equation}
v_{2}(\eta)/ \epsilon,
\end{equation},
with \begin{equation} \epsilon=
\frac{\sqrt{1+\beta}-\sqrt{1-\beta}}{\sqrt{1+\beta}}
\end{equation}
 having the limits
\begin{equation}
\beta \rightarrow 0 \mbox{, } \epsilon \rightarrow \beta
\rightarrow 0
\end{equation}
and
\begin{equation}
\beta\rightarrow 1 \mbox{ , } \epsilon \rightarrow 1
\end{equation}
We see that

    $i)$ $N_{A}\rightarrow A$, $R\rightarrow 0$,
    $v_{2}(\eta)/ \epsilon \rightarrow$ constant, increasing with
$\sqrt{s}$ and $N_{A}$

    $ii)$ $N_{A}\rightarrow 0$,  $\beta\rightarrow 1$,
 $v_{2}(\eta)/\epsilon \rightarrow 0$.

In order to compare with experimental data the dependence of
$v_{2}$ on the pseudorapidity, we start with the
$\frac{dn}{d\eta}$ data of PHOBOS collaboration [20] taken at
$N_{part}=211$. From formula (12) we compute $\eta^{t}$ at each
value of $\eta$ and then $v_{2}$ using equation (31). Our result
together with the experimental data [20] is presented in fig 1. In
the same way, using equation (31) we compute the dependence of
$v_{2}$ on the number of participants. In fig 2. we show our
results together with the experimental data. In both cases,
rapidity and centrality dependence, the agreement is very good.

Summarizing up, the analytical formulae (21) and (34) obtained in
the framework of string percolation are able to describe rightly
the dependence of the elliptical flow on rapidity and centrality.

\begin{figure}
\begin{center}
    \begin{tabular}{cc}
     \resizebox{80mm}{!}{\includegraphics{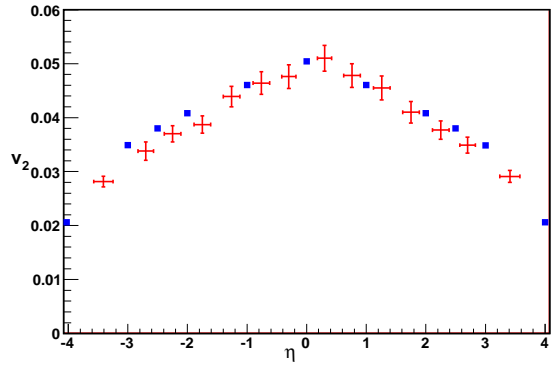}}
    \end{tabular}
    \caption{Elliptic flow as a function of pseudorapidity for
$N_{part}=211$ in Au+Au collisions at energy $\sqrt{s}=$ 200 GeV.
Dots in blue are used for our results and bars in red are data
taken from reference [20].}
    \label{test4}
\end{center}
\end{figure}
\begin{figure}
\begin{center}
    \begin{tabular}{cc}
     \resizebox{80mm}{!}{\includegraphics{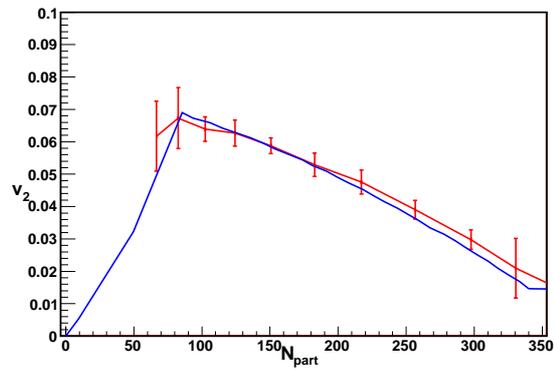}}
    \end{tabular}
    \caption{Elliptic flow dependence on the number of participants, at energy $\sqrt{s}=$ 200 GeV. Results compared to
PHOBOS data. Lines in blue are used for our results and red lines are data
taken form reference [20].}
    \label{test4}
\end{center}
\end{figure}
\section{Acknowledgements}
This work is under the project FPA2008-01177 of Spain, and of the
Xunta de Galicia. J. Dias de Deus thanks the support of the
FCT/Portugal project PPCDT/FIS/575682004.


\begin{thebibliography}{12}
\medskip
\bibitem{Adcox:2004mh}
  K.~Adcox {\it et al.}  [PHENIX Collaboration],
  Nucl.\ Phys.\  A {\bf 757} (2005) 184.
\bibitem{Adams:2005dq}
  J.~Adams {\it et al.}  [STAR Collaboration],
  Nucl.\ Phys.\  A {\bf 757}, 102 (2005).
  C.~Adler {\it et al.}  [STAR Collaboration],
  Phys.\ Rev.\ Lett.\  {\bf 87}, 182301 (2001).
\bibitem{Manly:2005zy}
  S.~Manly {\it et al.}  [PHOBOS Collaboration],
  Nucl.\ Phys.\  A {\bf 774}, 523 (2006).
\bibitem{Borghini:2007ub}
  N.~Borghini and U.~A.~Wiedemann,
  J.\ Phys.\ G {\bf 35}, 023001 (2008).
\bibitem{Ollitrault:1992bk}
  J.~Y.~Ollitrault,
  Phys.\ Rev.\  D {\bf 46}, 229 (1992).
\bibitem{Huovinen:2001cy}
  P.~Huovinen, P.~F.~Kolb, U.~W.~Heinz, P.~V.~Ruuskanen and S.~A.~Voloshin,
  Phys.\ Lett.\  B {\bf 503}, 58 (2001).
\bibitem{Bravina:2004td}
  L.~Bravina, K.~Tywoniuk, E.~Zabrodin, G.~Burau, J.~Bleibel, C.~Fuchs and A.~Faessler,
  Phys.\ Lett.\  B {\bf 631}, 109 (2005).
\bibitem{Teaney:2000cw}
  D.~Teaney, J.~Lauret and E.~V.~Shuryak,
  Phys.\ Rev.\ Lett.\  {\bf 86}, 4783 (2001).
  T.~Hirano, U.~W.~Heinz, D.~Kharzeev, R.~Lacey and Y.~Nara,
  Phys.\ Rev.\  C {\bf 77}, 044909 (2008).
\bibitem{Molnar:2001nk}
  D.~Molnar and M.~Gyulassy,
  Nucl.\ Phys.\  A {\bf 698}, 379 (2002).
\bibitem{Bleicher:2007cs}
  M.~Bleicher and X.~Zhu,
  Eur.\ Phys.\ J.\  C {\bf 49}, 303 (2007).
\bibitem{Lin:2002cs}
 Lin Z W and Ko C M  Phys. Rev. C 65 034904, (2002);
Chen L W and Ko C M Phys. Lett. B 634 205,(2006)
\bibitem{Bautista:2009my}
  I.~Bautista, L.~Cunqueiro, J.~D.~de Deus and C.~Pajares,
  J.\ Phys.\ G {\bf 37} (2010) 015103
\bibitem{Torrieri:2009fv}
  G.~Torrieri,
  arXiv:0911.4775 [nucl-th].
\bibitem{Pajares:2005kk}
  C.~Pajares,
  Eur.\ Phys.\ J.\  C {\bf 43}, 9 (2005)
  J.~Dias de Deus and R.~Ugoccioni,
  Eur.\ Phys.\ J.\  C {\bf 43}, 249 (2005).
\bibitem{Armesto:1996kt}
  N.~Armesto, M.~A.~Braun, E.~G.~Ferreiro and C.~Pajares,
  Phys.\ Rev.\ Lett.\  {\bf 77}, 3736 (1996)
  M.~A.~Braun and C.~Pajares,
  Phys.\ Rev.\ Lett.\  {\bf 85}, 4864 (2000)
M.~A.~Braun, F.~Del Moral and C.~Pajares,
  Phys.\ Rev.\  C {\bf 65}, 024907 (2002)
\bibitem{McLerran:2001cv}
  L.~D.~McLerran and J.~Schaffner-Bielich,
  Phys.\ Lett.\  B {\bf 514}, 29 (2001)
  J.~Schaffner-Bielich, D.~Kharzeev, L.~D.~McLerran and R.~Venugopalan,
  Nucl.\ Phys.\  A {\bf 705}, 494 (2002)
\bibitem{DiasdeDeus:2003fg}
  J.~Dias de Deus, E.~G.~Ferreiro, C.~Pajares and R.~Ugoccioni,
  Phys.\ Lett.\  B {\bf 581}, 156 (2004)
\bibitem{DiasdeDeus:2003ei}
  J.~Dias de Deus, E.~G.~Ferreiro, C.~Pajares and R.~Ugoccioni,
  Eur.\ Phys.\ J.\  C {\bf 40}, 229 (2005)
\bibitem{Cunqueiro:2007fn}
  L.~Cunqueiro, J.~Dias de Deus, E.~G.~Ferreiro and C.~Pajares,
  Eur.\ Phys.\ J.\  C {\bf 53}, 585 (2008)
\bibitem{Back:2004mh}
  B.~B.~Back {\it et al.}  [PHOBOS Collaboration],
  Phys.\ Rev.\  C {\bf 72}, 051901 (2005)
 B.~B.~Back {\it et al.}  [PHOBOS Collaboration],
  Phys.\ Rev.\ Lett.\  {\bf 97} (2006) 012301


\end{thebibliography}
\end{document}